\documentclass[twocolumn]{aastex61}

\hypersetup{linkcolor=red,citecolor=blue,filecolor=cyan,urlcolor=magenta}

\usepackage[all]{hypcap}
\usepackage{amsmath}
\usepackage{aas_macros}
\usepackage{graphics}
\usepackage{amsmath}
\usepackage{natbib}
\newcommand{\myemail}{adams@psi.edu}
\newcommand{\kepler}{\emph{Kepler}}
\newcommand{\ktwo}{\emph{K2}}

\received{November 1, 2016}
\revised{\today}
\submitjournal{AAS}

\begin{document}


\title{Ultra Short Period Planets in \ktwo\ with companions: a double transiting system for EPIC 220674823}


\author{Elisabeth R. Adams}
\affil{Planetary Science Institute, 1700 E. Ft. Lowell, Suite 106, Tucson, AZ 85719, USA}
\email{\myemail}

\author{Brian Jackson}
\affil{Department of Physics, Boise State University, 1910 University Drive, Boise ID 83725, USA}

\author{Michael Endl}
\affil{McDonald Observatory, The University of Texas at Austin, Austin, TX 78712, USA}

\author{William D. Cochran}
\affil{McDonald Observatory, The University of Texas at Austin, Austin, TX 78712, USA}

\author{Phillip J. MacQueen}
\affil{McDonald Observatory, The University of Texas at Austin, Austin, TX 78712, USA}

\author{Dmitry A. Duev}
\affil{California Institute of Technology, Pasadena, CA, 91125, USA}

\author{Rebecca Jensen-Clem}
\affil{California Institute of Technology, Pasadena, CA, 91125, USA}

\author{Ma{\"i}ssa Salama }
\affil{Institute for Astronomy, University of Hawai`i at M\={a}noa, Hilo, HI 96720-2700, USA}

\author{Carl Ziegler}
\affil{University of North Carolina at Chapel Hill, Chapel Hill, NC 27599, USA}

\author{Christoph Baranec}
\affil{Institute for Astronomy, University of Hawai`i at M\={a}noa, Hilo, HI 96720-2700, USA}
 
\author{Shrinivas Kulkarni}
\affil{California Institute of Technology, Pasadena, CA, 91125, USA}

\author{Nicholas M. Law}
\affil{University of North Carolina at Chapel Hill, Chapel Hill, NC 27599, USA}

\author{Reed Riddle}
\affil{California Institute of Technology, Pasadena, CA, 91125, USA}


\begin{abstract}

Two transiting planets have been identified orbiting K2 target EPIC 220674823. One object is an ultra-short-period planet (USP) with a period of just 0.57 days (13.7 hours), while the other has a period of 13.3 days. Both planets are small, with the former having a radius of $R_{\rm p1}=1.5~R_{\oplus}$ and the latter $R_{\rm p2}=2.5~R_{\oplus}$.  Follow-up observations, including radial velocity (with uncertainties of 110 m$s^{-1}$) and high-resolution adaptive optics imagery, show no signs of stellar companions. EPIC 220674823 is the 12th confirmed or validated planetary system in which an ultra-short-period planet (i.e., having an orbital period less than 1 day) is accompanied by at least one additional planet, suggesting that such systems may be common and must be accounted for in models for the formation and evolution of such extreme systems.

\end{abstract}


\section{Introduction}
\label{sec:Introduction}

With orbital periods of less than 1 day, ultra-short-period planets (USPs) and candidates represent a relatively rare class of planet, orbiting about 0.1\% of Sun-like stars \citep{SanchisOjeda2014}. However, they also represent an enormous opportunity for observational follow-up since their proximity to their host stars means that they are more likely to induce measurable radial velocity (RV) signals than similarly massed planets farther out, and are also more likely to transit their host stars. In addition, they present an unexploited opportunity to learn about planet formation and the early evolution of planetary systems since they probably could not form where we find them today and instead may require migration \citep{2013ApJ...779..165J}. 

In several cases, USPs are observed with sibling planets on longer-period orbits, and \citet{SanchisOjeda2014} proposed that even those USP systems without known siblings likely have nontransiting additional planets. These results suggest that perhaps the origins of USPs involve interactions among the sibling planets. For instance, in considering the evolution of 55 Cnc e, an 8 $M_{\oplus}$ (Earth mass) planet with $P=0.7$ days in a five-planet system, \citet{2015MNRAS.450.4505H} argued that secular resonances in the system may have excited the planet's orbital eccentricity, which, coupled with tidal interactions with the host star, probably drove substantial inward migration. Because the periods of the innermost planets are so short, it is common for them to be separated by tens of Hill radii from the next planets out in the system. Moreover, no USP has been identified in a mean motion resonance with other known planets, although a few are close -- the period ratio of Kepler-32 f:e is 3.9, while that of Kepler-80 f:d is 3.1. 

Tidal evolution powered by multiplanet interactions may be required to move USPs into their current orbits since many USPs currently occupy space that was once inside the stellar surface \citep{2016CeMDA.126..227J}, and they are too small to raise a substantial tide within their host star. Orbital decay powered by tidal dissipation within the host star would take more than 100 Gyr to move Kepler-78 b, with a mass $M_p =1.7 M_{\oplus}$ \citep{2013Natur.503..381H, 2013Natur.503..377P}, from $P=1$ day to its current orbit of $P = 8.5$ hr (assuming a tidal dissipation efficiency for the host star of $Q_\star = 10^7$; \citealp{2012ApJ...751...96P}). 

Instead of tidal decay powered by secular or resonant interactions, another possibility is that USPs arrived in their orbits as the result of dynamical excitation or scattering into an initial highly eccentric orbit by another planet or a stellar companion, followed by orbital circularization \citep{2007ApJ...669.1298F}. Such a scenario is statistically unlikely since there is a narrow range of initial orbits with pericenters small enough for tidal circularization to be important but large enough that the proto-USP does not plunge into the host star -- though perhaps the small occurrence rate of USPs is consistent with that low probability. In either case, if multiplanet interactions, whether through tidal migration or through dynamical excitation, are required to emplace USPs at their current location, then the presence of an USP may be the signpost of additional, unseen planets in a system. The sizes and orbits of any additional planets will also help discriminate between origin scenarios.

Of special interest are the systems discovered by \ktwo, the host stars for which tend to be brighter than stars monitored by the \kepler\ mission, making them far more amenable to follow-up characterization of the host stars and planets. Here we report on a multiplanet system around EPIC 220674823, a $Kp=11.958$ star somewhat smaller than the Sun that hosts two transiting super-Earths at orbital periods of 0.57 and 13.3 days. These planets were detected in Campaign 8 (C08) data from the \ktwo\ mission as part of the ongoing efforts of the Short-Period Planets Group (SuPerPiG, \url{http://www.astrojack.com/research/superpig/}).

\begin{figure}
\includegraphics[width=0.5\textwidth]{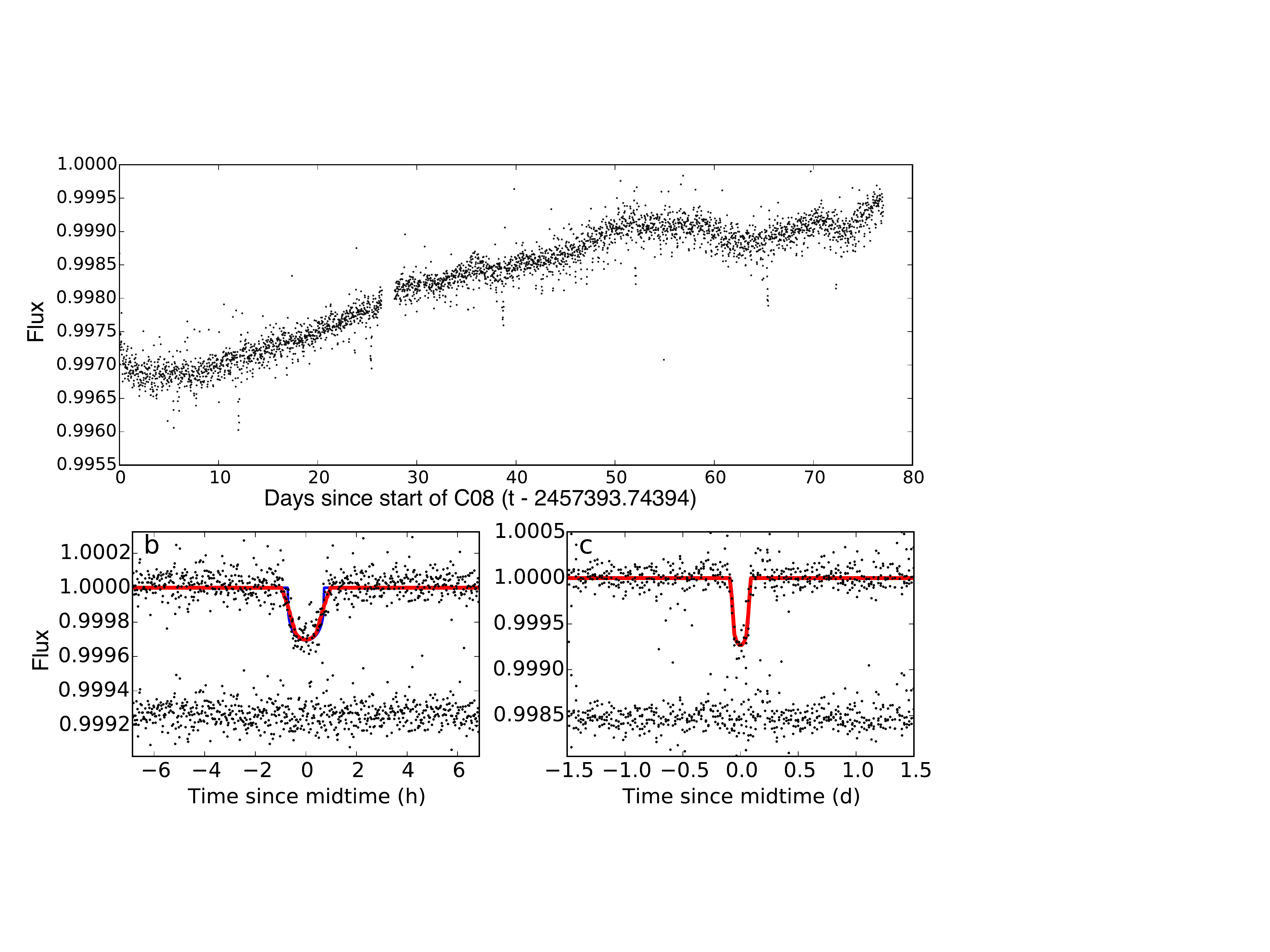}
\caption{Top panel: full photometric time series of EPIC 220674823 from k2sff via MAST \citep{2014PASP..126..948V}, with transits by planets p1 and p2 clearly visible. Bottom panels: best-fit light curves to planets p1 (b) and p2 (c). Light curves are folded to the orbital period and truncated around $\pm1.5$~ days from mid-transit for p2 in (c), and they have been binned for presentation.}
\label{fig:transits}
\end{figure}

\section{Photometry and Transit Search}
\label{section:planet}

In this section, we describe our data conditioning and search process. The innermost planet, p1, was found using the same search method described in more detail in our analysis of Campaigns 0-5 \citep{2016AJ....152...47A}. Briefly, we retrieved the k2sff photometry \citep{2014PASP..126..948V} for C08 and applied a median boxcar filter with a width of 1 day. We masked out 10$\sigma$ outliers and searched for periodic signals between 3 and 24 hr using the EEBLS algorithm \citep{2002A&A...391..369K}. EPIC 220674823 was among the candidates identified. We then fit the transit light curve using the algorithm from \citet{2002ApJ...580L.171M}, as implemented by the Python package Batman \citep{2015PASP..127.1161K}. We fixed eccentricity at zero for both planets; a separate test for p2 found that the eccentricity value was not well constrained by the light curve.  We used the pymodelfit and PyMC packages to conduct a Markov chain Monte Carlo (MCMC) transit analysis using 100,000 iterations (discarding as burn-in the first 1000 iterations and thinning the sample by a factor of 10). The resulting best-fit model parameters are shown in \autoref{table:cands}, with the quoted 1$\sigma$ error bars containing 68.3\% of the posterior values. 

\begin{figure}
\includegraphics[width=0.5\textwidth]{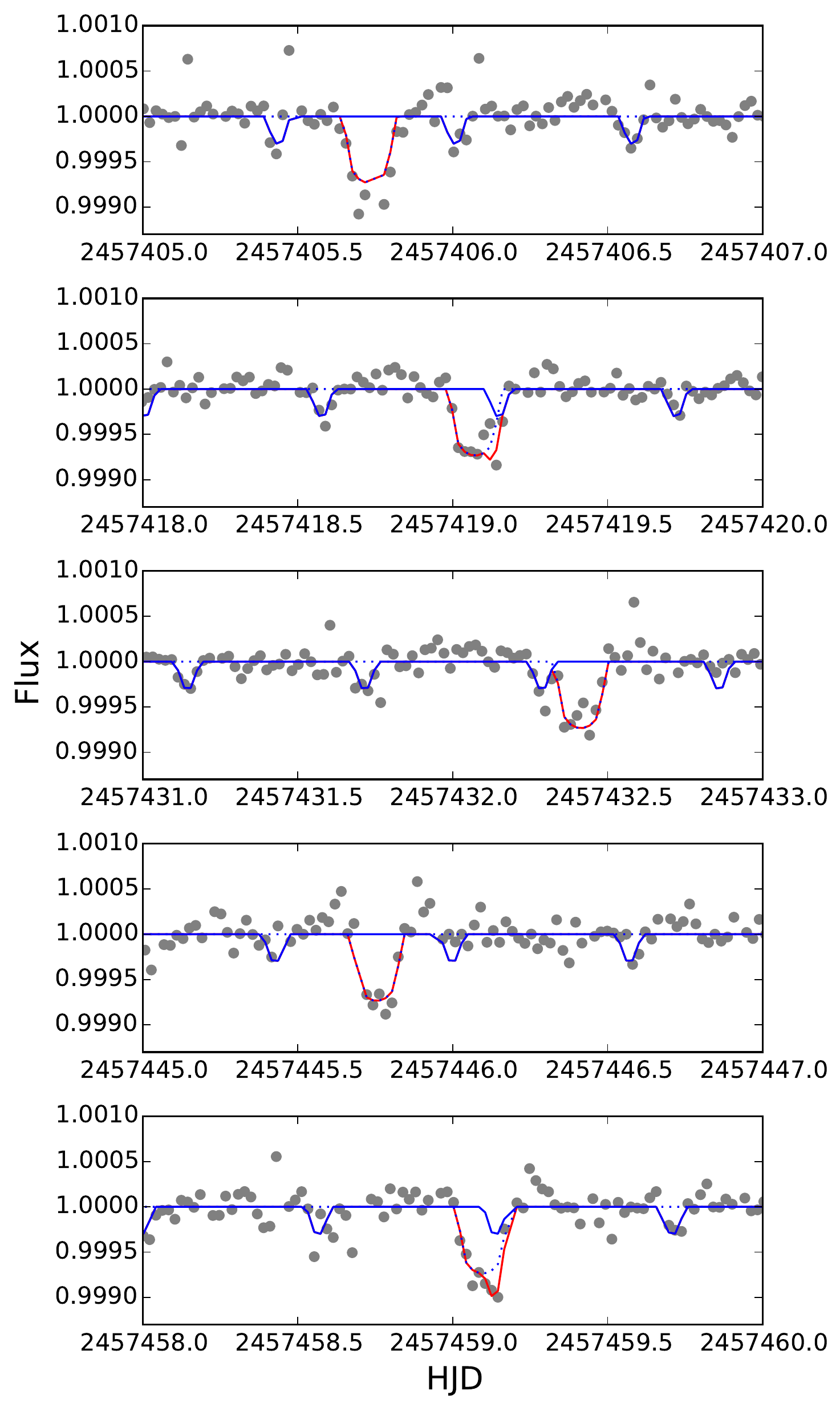}
\caption{All five transits of planet p2. The blue solid line shows the fit to the inner planet, p1 ($P=13$ hr), while the blue dotted line shows the fit to the outer planet, p2 ($P=13$ days). In three of the five transits of p2 the light curves overlap, so the combined model plot is shown in red.}
\label{fig:p2}
\end{figure}

After p1 was identified, we examined an unfolded time series of the raw flux for the star for the entire campaign. This revealed five clear transits of an additional larger object, p2, with a period of about 13 days (see \autoref{fig:transits}). Since several of the transits of the larger planet overlap with transits of p1 (\autoref{fig:p2}), we modified our light-curve fitting process as follows. We first subtracted out the best-fit light curve for p1 from the photometric series and applied the EEBLS transit search scheme \citep{2002A&A...391..369K} to the remaining signal to estimate p2's orbital period in the same manner as the search for p1, except centered around $P=13\pm1$ days. In this way we refined the orbital period estimate to $P_{p2} = 13.341245\pm0.0001$. We then folded the photometry around this period and fit for the best model of p2. For consistency, we took this model for p2 and subtracted it from the original photometry, so that both p1 and p2 would be fit using time series that excluded the other planet. We then reran the fits on p1 using the subtracted photometry in the same manner as before. The results for p1 were almost identical either way (not surprising, since only three out of hundreds of transits overlapped), but for p2 the difference was important (three out of five transits overlapped). The parameters shown in \autoref{table:cands} are from fits made to data where the other planet's transit model has been subtracted out.

As a check on stellar variability, we took the raw time series and excluded the times during both planetary transits to make a Lomb-Scargle periodogram. The only peaks of significance were around 48.9 days (possibly related to the stellar rotation period) and less than 15 minutes (likely stellar oscillations).

\capstartfalse
\begin{deluxetable*}{lll}
\tablewidth{0pt}
\tablecaption{EPIC 220674823 System Parameters}
\tablehead{
Planetary Parameters &  b (p1) & c (p2)
}
\startdata
Period (days)			&$0.571308\pm0.00003$				   	&$13.341245\pm0.0001$ \\
$T_0$ 		     		&$2457437.42948_{-0.00085}^{+0.00086}$	&$2457405.73124_{-0.00386}^{+0.0027}$ \\
$R_{\rm p}/R_\star$		&$0.0161_{-0.0006}^{+0.0013}$			&$0.028_{-0.0008}^{+0.0004}$ \\
$R_{\oplus}$			&$1.46\pm0.14$						&$2.53\pm0.14$ \\
$a/R_\star$			&$2.8_{-0.6}^{+0.3}$						&$16.6_{-13.2}^{+2.4}$\\
$i$					&$80.2_{-10.7}^{+7.0}$					&$87.2_{-13.7}^{+0.5}$ \\
$e$					& $0$ (fixed)							& $0$ (fixed) \\
$\sigma_{odd-even}$ 	&$0.1$								&$0.4$	\\
\hline
Stellar Parameters		& 						& Source \\
\hline
R.A. (deg)            	 	& $13.0797796$			& EPIC \\
Decl. (deg)				& $10.7946987$			& EPIC \\
Mag (Kep)				& $11.96$					& EPIC \\
Spectral type       	 	& \text{G5}				& Inferred from McDonald spectra and EPIC colors \\
Proper motion (mas)		& $60.3\pm3.7$			& Gaia DR1 \citep{2016AA...595A...1G, 2016AA...595A...2G} \\
Parallax (mas)			& $3.96\pm0.78$			& Gaia DR1 \citep{2016AA...595A...1G, 2016AA...595A...2G} \\
Inferred distance (pc)	& $253\pm50$				& Derived from Gaia parallax \\
Age	(Gyr)				& $5$					& Dartmouth isochrone \citep{2008ApJS..178...89D} \\
$R_\star$	($R_{\odot}$)	& $0.83\pm0.04$			& McDonald observations + \citet{Boyajian2012} \\
$M_\star$	($M_{\odot}$)	& $0.93\pm0.01$			& Dartmouth isochrone \citep{2008ApJS..178...89D} \\
Luminosity ($L_{\odot}$)	& $0.76$					& Derived from $M_\star$ \\
$u_1$				& $0.5078$				& Quadratic limb darkening from \citet{2011AA...529A..75C} \\
$u_2$				& $0.1615$				& Quadratic limb darkening from \citet{2011AA...529A..75C}  \\
\enddata
\tablecomments{Planetary parameters are from best MCMC fit with 68.3\% errors on the posterior distribution.}
\label{table:cands}
\end{deluxetable*}
\capstarttrue

\capstartfalse
\begin{deluxetable*}{ll lr ll lr}
\tablewidth{0pt}
\tablecaption{Spectra of EPIC 220674823}
\tablehead{
UT 			& HJD			 & $T_{\rm{eff}}$		&[Fe/H]			&log($g$)			& $v sin(i)$ 		& RV (km s$^{-1}$) 		&Notes
}
\startdata
2016 Aug 15  09:46:54  & 2457615.9038   & $5580\pm 86$	&$0.040\pm  0.03 $	&$ 4.62\pm0.16$	&$ 2.62\pm0.18$	&$-15.86\pm0.13$	&McDonald\\
2016 Sep 08  09:17:37  & 2457639.9016   & $5600\pm55$	&$0.010\pm 0.02$	&$ 4.50 \pm0.10$	&$ 2.23\pm0.23$	&$-16.01\pm0.10$	&McDonald \\
2016 Oct 11  07:27:13  & 2457672.8187   & $5660\pm60$       &$-0.010\pm 0.02$    &$ 4.62\pm0.07$	&$3.08\pm0.21$	&$-15.798 \pm0.28$	&McDonald \\
                              	  &				& $5590\pm51$	&$0.025\pm0.02$	&$4.56\pm0.09$	&				&				& Adopted average  \\
\hline
				  &			        & $5814\pm181$	&$-0.283\pm0.25$	& $4.409\pm0.085$	&				&				&EPIC \citep{Huber2016}\\
\enddata
\tablecomments{} 
\label{table:stars}
\end{deluxetable*}
\capstarttrue


\subsection{No timing variations}

\begin{figure}
\includegraphics[width=0.5\textwidth]{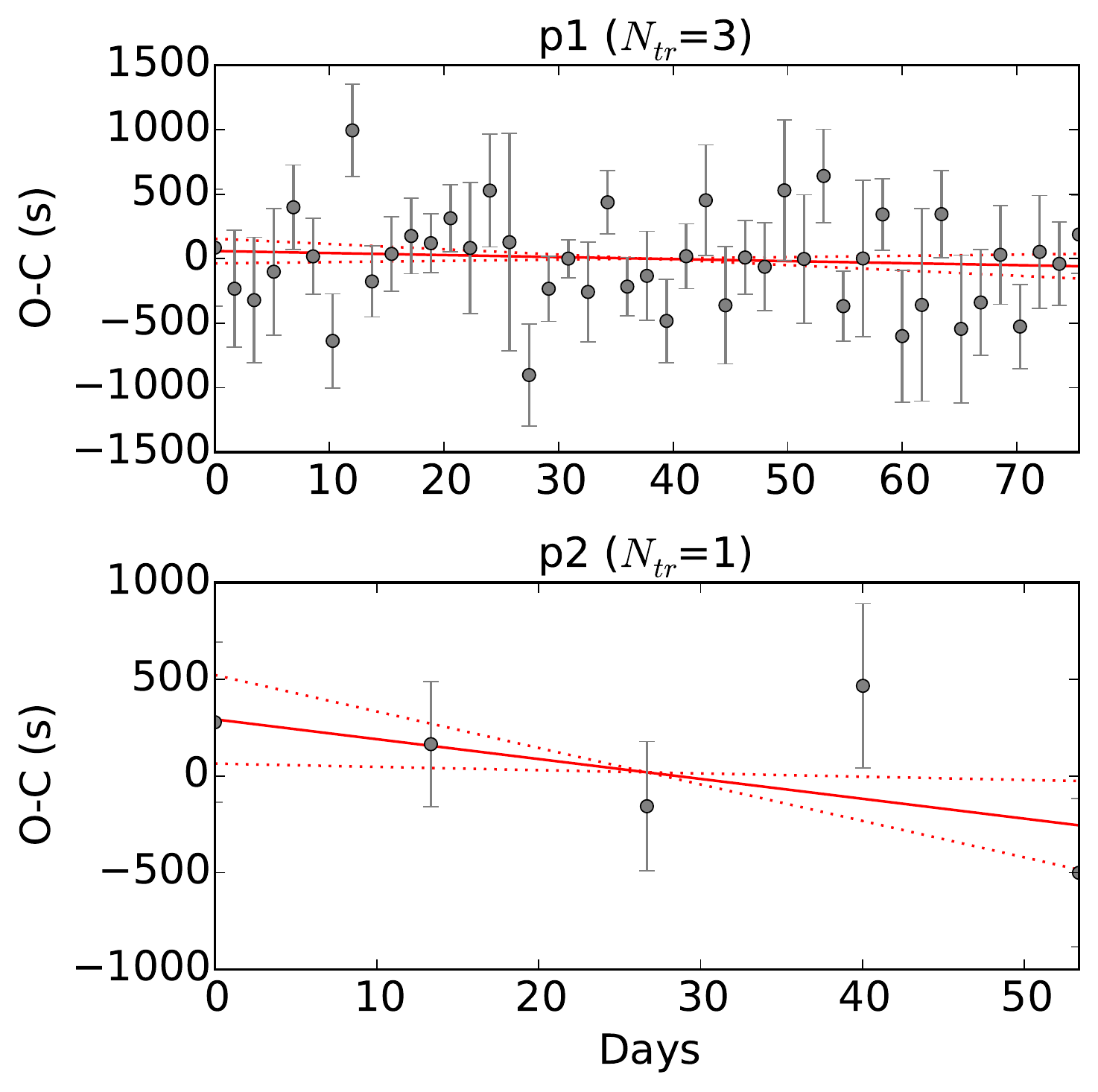}
\caption{Observed minus calculated mid-transit times for p1 (top) and p2 (bottom). For p1, which does not have sufficient precision to individually fit each transit, we stacked $N_{\rm tr}=3$ consecutive transits together, for one point every 1.7 days; the pattern is similar if more consecutive transits were stacked instead. For p2, each individual transit was fit.}
\label{fig:ominusc}
\end{figure}

As part of our analysis, we searched for transit timing variations in the fits to both planet candidates (\autoref{fig:ominusc}). The transit light curves for the smaller planet, p1, do not have sufficient signal-to-noise ratio (S/N) to allow us to robustly fit the mid-transit time of each transit. Instead, we folded together $N_{tr}=3$ consecutive transits to increase the S/N and fit a linear ephemeris to each stacked trio, giving one point for every 1.7 days \citep{2013ApJ...779..165J}. We also explored the effect of stacking together more consecutive transits and found similar fits for the ephemeris (modulo the loss of time resolution). For p2 each individual transit was fit. For both objects, the observed minus calculated ($O-C$) plots are consistent with no variation.

\section{Follow-up observations and validation}

Systems with multiple candidate transiting planets are much more likely than singleton systems to host genuine planets \citep{2012ApJ...750..112L}. However, follow-up observations are important for full validation and for determining the correct planetary parameters. We collected stellar spectra, low-resolution RV data, and high angular resolution imagery for this system and applied an open-source analysis package to statistically validate the objects as planets.

\subsection{Spectroscopic Observations}

\begin{figure}
\includegraphics[width=0.5\textwidth]{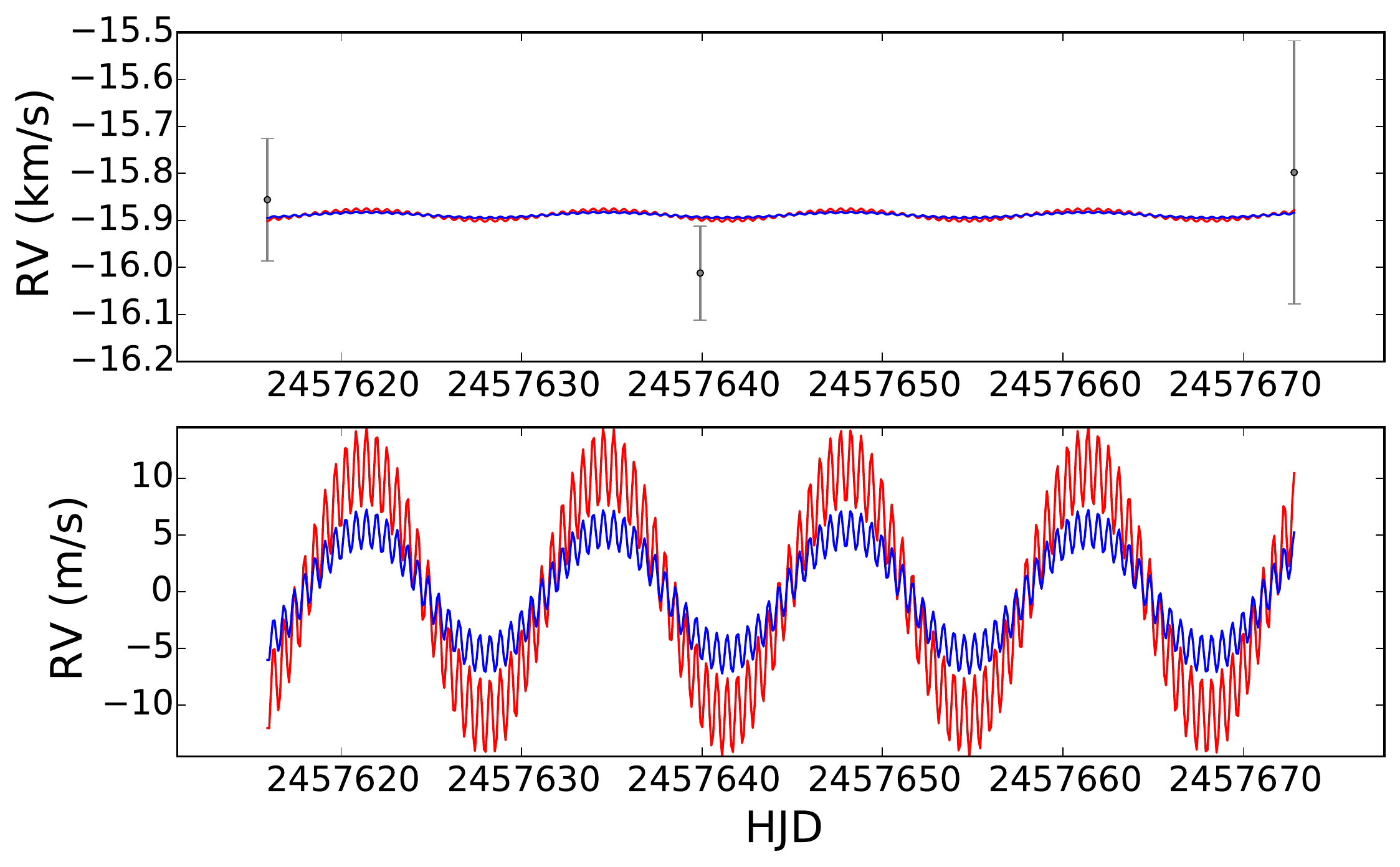}
\caption{Recon spectra of EPIC 220674823 show no stellar RV signal within errors ($RMS=110\ {\rm m\ s}^{-1}$). The combined estimated RV  signal of the two planets is plotted for reference (assuming masses of 4.5 and 41 $M_{\oplus}$ in blue and 2.3 and 20.5 $M_{\oplus}$ in red). The bottom panel shows a zoomed-in version of the estimated signal, which, with an estimated amplitude of 2-11 ${\rm m\ s}^{-1}$, is easily accessible by more precise RV instruments.}
\label{fig:rv}
\end{figure}

Reconnaissance spectra were obtained of EPIC 220674823 with the Tull Coud\'e spectrograph \citep{1995PASP..107..251T} at the Harlan J. Smith 2.7 m telescope at McDonald Observatory on three nights during 2016 August--October. The exposure times ranged from 600 to 1600 s, resulting in S/Ns from 30 to 60 per resolution element at 5650\AA. We determined stellar parameters for the host stars with the spectral fitting tool {\it Kea} \citep{2016PASP..128i4502E}, which compares high-resolution, low-S/N spectra of stars to a massive grid of synthetic stellar spectral models in order to determine the fundamental stellar parameters of the \kepler\ target stars. We also determined absolute RVs by cross-correlating the spectra with the RV-standard star HD~50692. The stellar parameters from each observation and our adopted values of  $T_{\rm eff}$, $log(g)$, and [Fe/H] are shown in \autoref{table:stars}. No radial velocity differences were seen within uncertainties (\autoref{fig:rv}). We used the observed $T_{\rm eff}$ and the models of \citet{Boyajian2012} to estimate the $R_\star$, and the stellar mass was estimated from those using the Dartmouth models and assuming a stellar age of 5 Gyr \citep{2008ApJS..178...89D}.

With an rms error of 110  ${\rm m\ s}^{-1}$, we can place an upper limit on the planets' masses of $M_{\rm p1} < 0.43 M_{\rm Jup}$ (Jupiter masses) 
and $M_{\rm p2} < 1.22\ M_{\rm Jup}$ 
using Equation 14 from \citet{2010exop.book...27L}. If we assume that the planets are entirely rocky, we can estimate the approximate masses using the mass-radius relation from \citet{2014ApJ...792....1L}: $M_{\rm p}/{\rm M_\odot} \approx \left( R_{\rm p}/{\rm R_\odot}\right)^4$, which yields masses of 4.5 and 41 Earth masses $M_{\oplus}$ for p1 and p2, respectively. The large radius for p2 suggests that it likely harbors a substantial gaseous envelope \citep{2015ApJ...801...41R}, and, following the mass-radius relationship described in \citet{2008ApJ...673.1160A}, would have a mass of about 20 $M_{\oplus}$, depending on the core composition and gas fraction. So we also provide an estimated RV signal if both planets had substantial volatiles, using half the core-only estimates, or 2.3 and 20.5 $M_{\oplus}$ for p1 and p2, respectively. The RV half-amplitude estimates are 3.5 and 11 ${\rm m}~{\rm s}^{-1}$ for p1 and p2, respectively, for the larger masses, and 1.7 and 5.5 ${\rm m}~{\rm s}^{-1}$ respectively, for the smaller masses, as shown in \autoref{fig:rv}. Even the smaller mass estimate is within reach of the best world-class instruments today. 

\subsection{Adaptive Optics Observations}

We observed EPIC 220674823 in $i'$ band with Robo-AO at Kitt Peak \citep{2014ApJ...790L...8B, 2016SPIE.9909E..1AS} on 2016 September 19 and 25. Each observation comprised a sequence of full-frame-transfer detector readouts of an electron-multiplying CCD camera at the maximum rate of 8.6 Hz for a total exposure time of 90 s. Individual frames are dark- and flat-field corrected before being registered to correct for the dynamic image displacement of the target that cannot be measured with the laser guide star, and co-added. We detected no stellar companions within 2 mag at 0\farcs2, nor within 4.5 mag at 1\farcs0 of the primary target. 

\begin{figure}
\includegraphics[width=0.5 \textwidth]{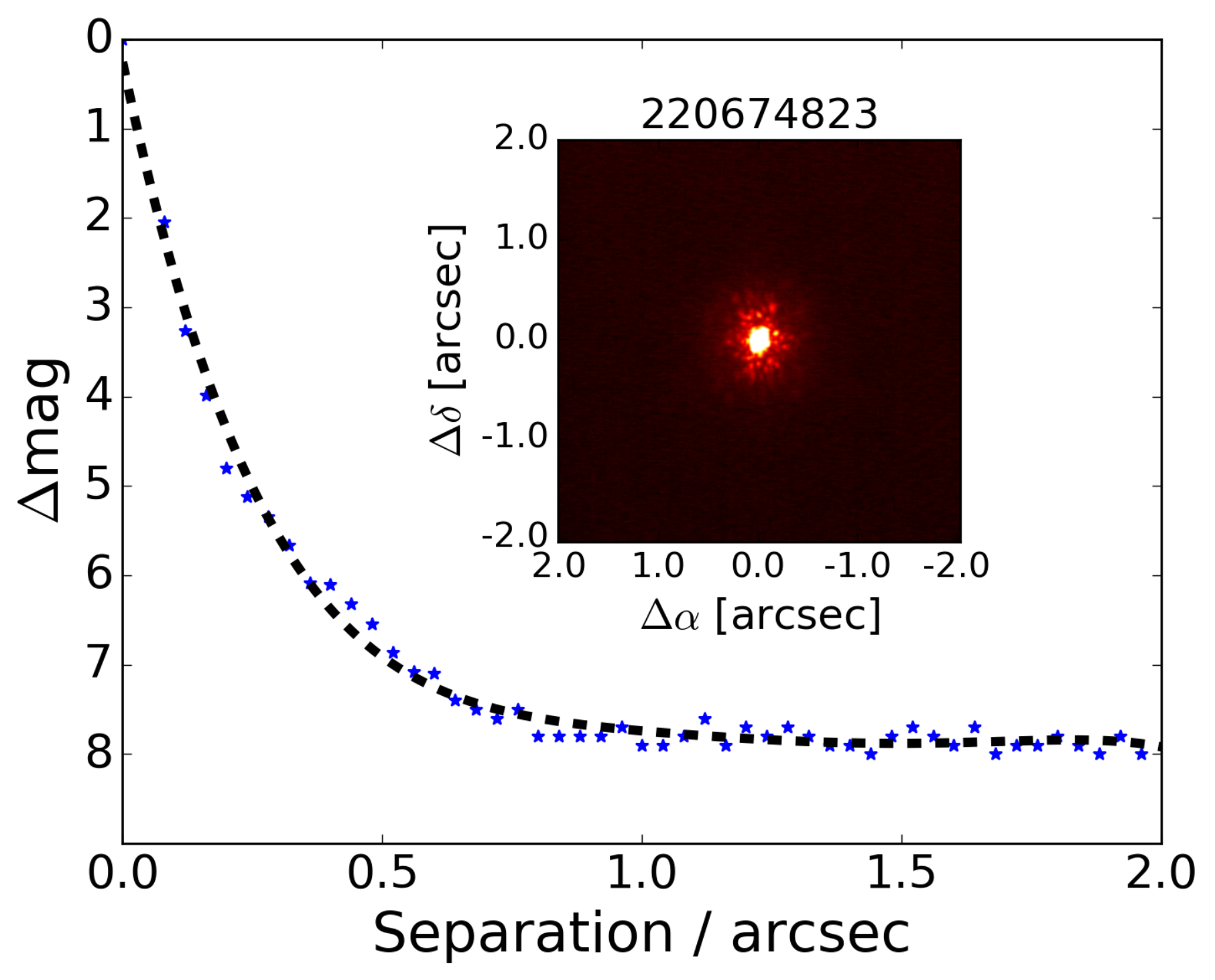}
\caption{The AO image of EPIC 220674823 in $K$ band with Keck/NIRC2 revealed no additional stars. Blue stars in the sensitivity curve are the measured minimal brightness of a possible companion consistent with a 5$\sigma$ detection; the black dashed line is a fitting function).}
\label{fig:keck}
\end{figure}

We made additional observations with the NIRC2 camera behind the Keck II adaptive optics system in natural guide star mode on 2016 October 16. We obtained six frames, 2 $\times$ 15 s exposures at each of three different dither positions in the Kp filter. After sky subtraction and flat-field calibration, the frames were co-added into a single image based on the automatic detection of the location of the target star in each dither frame. Using the same methodology of \citet{2016arXiv160503584Z}, a custom locally optimized point spread function was subtracted from the image, which was run through an automatic companion detection pipeline.  This allowed even tighter constraints to be placed: there are no companions as faint as $\Delta M = 2$ at a separation of 0.08\arcsec\ and $\Delta M = 8$ at 1\arcsec\ and beyond. A plot of the full Keck detection constraints is shown in ~\autoref{fig:keck}. A full description of the methodology for the automated search and the generation of the contrast curve can be found in Sections 3.5 and 3.6 of \citet{2016arXiv160503584Z}.

\subsection{Validation}
We used the freely available \emph{vespa} package to validate the planetary system \citep{2012ApJ...761....6M, 2015ascl.soft03011M}. To place a limit on secondary events, we subtracted the transit model of both planets from the photometry and then searched for the strongest periodic signal within 0.001 days of the planetary orbital period using EEBLS. This placed an upper limit on secondary events of 30 and 80 parts per million (ppm) for p1 and p2, respectively. Using the nondetection of secondaries, publicly available photometry from the ExoFOP\footnote{\url{https://exofop.ipac.caltech.edu/k2/edit_target.php?id=220674823}}, and our observational constraints, \emph{vespa} returned a false-positive probability (FPP) of $4\times10^{-5}$ for p1 and $6\times10^{-6}$ for p2, with each planet considered separately. Since they are part of a multitransiting system, the odds of a false positive are even lower \citep{2012ApJ...750..112L}. Thus, we consider EPIC 220674823 p1 and p2 to be validated and suggest that they be referred to as EPIC 220674823 b and EPIC 220674823 c, respectively.

\section{Discussion}

\subsection{Additional Planets in Systems with USPs}

\begin{figure}
\includegraphics[width=\textwidth]{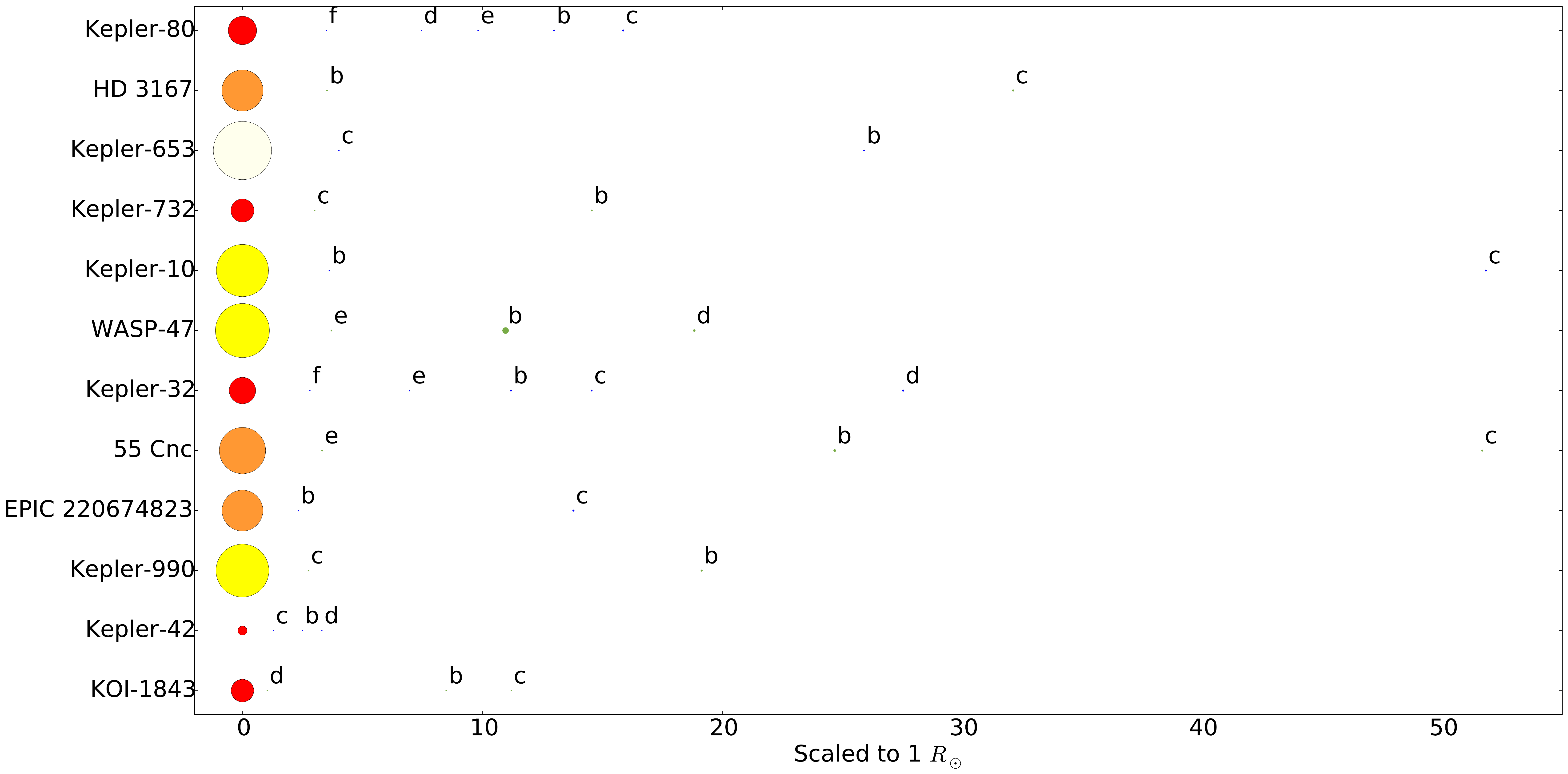}
\caption{Twelve multiplanet systems where the innermost member is a USP. All planet and star sizes and separations are plotted to scale ($R_{\odot}=1$). The planetary systems are sorted by orbital period of the innermost planet, with KOI-1843 d at 4 hr and Kepler-80 f at just under a day. The planet colors alternate systems between green and blue, and stars are color-coded by approximate spectral types: M (red, $R_\star < 0.7$), K (orange, $0.7 < R_\star \le 0.95$), G (yellow, $0.95 < R_\star \le 1.15$), and F (off-white, $1.15 < R_\star \le 1.4$).}
\label{fig:multi_systems}
\end{figure}

We have identified another USP that is part of a multiplanet system. We know of at least 12 confirmed systems of USPs with one to four additional planets each (see \autoref{fig:multi_systems}), including the new EPIC 220674823 system. While USPs have high geometric transit probabilities (10-50\%), the transit probabilities for their sibling planets, with orbital periods of 1-45 days, drop below a few percent. In 10 of the 12 systems, all of the known planets transit. For the WASP-47 system, two of the three companions transit, and the outermost planet was identified in RV only. The remaining system, 55 Cnc, was discovered via RV observations, with the innermost planet, 55 Cnc e, originally identified at the wrong period of $P=2.8$ days \citep{2004ApJ...614L..81M}. Once the true period of less than 1 day was recognized, the transit probability for 55 Cnc e was revised upward to 25\% \citep{2010ApJ...722..937D}, and its transit was observed \citep{2011A&A...533A.114D, 2011ApJ...737L..18W}, although no other planet in the system has been shown to transit to date.

Are USPs statistically more likely to exist in multiplanet systems than as singletons? We can estimate the likelihood with a simple heuristic argument. If we take the 11 USPs with at least one known transiting companion, we find that the sibling planet nearest the USP has a period $P=1.2-45$ days and a transit detection probability of 2-10\%. If we assume that every USP has at least one companion (transiting or not) that falls within the same orbital period range, then for the 175 USP systems listed at \url{exoplanets.org} in 2016 October, we would expect to have discovered $9.8\pm0.3$ companions (based on 10,000 random draws from the geometric probabilities of the known planets). This is nearly the same as the known number of 11 multiplanet USP systems. Thus, the current rate of detection of multiplanet systems is consistent with every USP having at least one additional companion within $P<45$ days. Note that half of the USP multi-systems have one additional companion, while the other half have two to four planets, but we have only considered the odds of the first, closest companion.

The argument above makes no assumption about whether planets preferentially have aligned orbital planes, which has the known effect of increasing the transit probabilities for the additional planetary companions. In the case of the triple system Kepler-18, \citet{2011ApJS..197....7C} showed that assuming that the mutual inclination of the planets is less than a few degrees increases the transit probability for the second planet by a factor of 2. We can incorporate into our heuristic calculation above the assumption that the next-nearest planet to a USP has an orbital inclination within a few degrees of the USP. This assumption roughly doubles the transit probabilities, resulting in an expected $19.6\pm0.6$ detections (instead of the actual 11 detected). Thus, if the next-nearest planets in USP systems are nearly coplanar to the USP, we find that between 47\% and 66\% of USP systems should host at least one additional planet, rather than 100\%. (Note that none of these arguments account for biased detection efficiencies.)


The truth probably lies between these two scenarios, since coplanarity cannot be assumed for all systems. Substantial misalignments have been reported in the literature: the USP 55 Cnc e has an inclination of $i_e=83.4^{\circ}$, while astrometric measurements of the outermost planet, 55 Cnc d, found that $i_d=53^{\circ}$ \citep{2004ApJ...614L..81M}. Some origin theories, particularly those invoking scattering, produce wide ranges of inclinations. Ultimately, the number of planets in multiple systems and the types of orbits the companion planets occupy will be a powerful constraint on planet formation theories.

\subsection{Small Planets in USP systems}

Among the 24 known companion planets to USPs in 12 systems, 21 are small, ranging from sub-Earths to Neptune-sized, or $R_{\rm p}=0.57-3.6\ R_\oplus$ (Kepler-42 d and WASP-47 d, respectively), with a median value of $R_{\rm p}=2.3\ R_\oplus$. The 12 USPs themselves are much smaller than their more distant companions, with an average radius of $1.3\ R_\oplus$, and a range of radius values from $R_{\rm p}=0.59-2.1\ R_\oplus$ (KOI-1843 d and 55 Cnc e, respectively). This could indicate the difference between the extreme environment of sun-grazing worlds, which have been stripped of their gases, and orbits just outside the most extreme zone. 

Only three companion planets have larger radius values: the RV planets 55 Cnc d ($P=5169$ days) and 55 Cnc b ($P=14.7$ days), which do not transit, but have sufficiently high masses (3.8 and 0.8 $M_{\rm Jup}$, respectively) that standard mass-radius relationships predict radii substantially larger than that of Neptune, and the transiting Hot Jupiter WASP-47 b ($P=4.2$~d), which has $R_{\rm p} = 1.15$ $R_{\rm Jup}$. Both of these systems are atypical in that (1) they are the only two systems that were identified from the ground, rather than through \kepler\ or \ktwo\ photometry, and (2) all of the planets in these two systems are larger than the planets in the other ten multiplanet systems. We stress that there may well be more distant companions like 55 Cnc d around other systems, since most USP-hosting stars have not been systematically monitored for long-term RV signals. 

The upshot of all these considerations is that the known planets in systems that host USPs tend to be smaller than Neptune. This tendency is unlikely to be due to observational biases since they would skew detections toward larger planets.

\subsection{USP-hosting stars Are High-Priority Candidates for Follow-up}

Despite their modest masses, USPs should be high-priority targets for precision RV observations, since the RV amplitude is a function of orbital distance. To date, the smallest planet measured with RV is Kepler-78 b, a USP with $P=0.35$ days and a mass of $M=1.69-1.85 M_{\oplus}$ \citep{2013Natur.503..377P, 2013Natur.503..381H}. Another advantage is that observations made over a few nights will span several orbital periods, allowing detections on a short timescale. The orbital periods of these objects are also well separated from the typical stellar variability period, decreasing potential confusion \citep{2011A&A...528A...4B}. Detecting the more distant companion planets in systems with USPs may also be possible with longer-term monitoring and could test theories of USP origins.

An advantage of \ktwo\ is that the average planet candidate is several magnitudes brighter than the typical \kepler\ target, putting them within easier reach of RV measurements. Among the multiplanet systems, the USP in three of the brightest five systems was first identified with \ktwo\ (HD 3167, $V=8.9$; WASP-47, $V=11.0$; and EPIC 220674823, $V=11.958$), with \kepler\ responsible for one (Kepler-10, $V=11.157$) and the last identified with RV from the ground because its star was so bright (55 Cnc, $V=5.95$).

Understanding the USP population has implications for the \emph{Transiting Exoplanet Survey Satellite (TESS)}, scheduled to launch in 2017, which will look for short-period rocky planets around $\sim$ 500,000 nearby, bright stars. Roughly 0.1\% of Sun-like stars host USPs, so \emph{TESS} should find hundreds of bright USPs. These planets would be ideal for follow-up, and a clear framework for their origins would motivate and guide additional \emph{TESS} observations.

\acknowledgments

Some of the data analyzed in this paper were collected by the \ktwo\ mission, funding for which is provided by the NASA Science Mission Directorate. The data were obtained from the Mikulski Archive for Space Telescopes (MAST). STScI is operated by the Association of Universities for Research in Astronomy, Inc., under NASA contract NAS5-26555. Support for MAST for non-HST data is provided by the NASA Office of Space Science via grant NNX09AF08G and by other grants and contracts. 
Some of the data presented herein were obtained at the W.M. Keck Observatory, which is operated as a scientific partnership among the California Institute of Technology, the University of California, and the National Aeronautics and Space Administration. The Observatory was made possible by the generous financial support of the W. M. Keck Foundation. The authors wish to recognize and acknowledge the very significant cultural role and reverence that the summit of Maunakea has always had within the indigenous Hawaiian community.  We are most fortunate to have the opportunity to conduct observations from this mountain. 
Robo-AO KP is a partnership between the California Institute of Technology, the University of Hawai`i, the University of North Carolina at Chapel Hill, the Inter-University Centre for Astronomy and Astrophysics, and the National Central University, Taiwan. Robo-AO KP is supported by a grant from Sudha Murty, Narayan Murthy, and Rohan Murty, and by a grant from the John Templeton Foundation. The Robo-AO instrument was developed with support from the National Science Foundation under grants AST-0906060, AST-0960343, and AST-1207891, from the Mt. Cuba Astronomical Foundation, and by a gift from Samuel Oschin.  Based (in part) on observations at Kitt Peak National Observatory, National Optical Astronomy Observatory (NOAO Prop. ID: 15B-3001), which is operated by the Association of Universities for Research in Astronomy (AURA) under cooperative agreement with the National Science Foundation. 
This work has made use of data from the European Space Agency (ESA) mission {\it Gaia} (\url{http://www.cosmos.esa.int/gaia}), processed by the {\it Gaia} Data Processing and Analysis Consortium (DPAC, \url{http://www.cosmos.esa.int/web/gaia/dpac/consortium}). Funding for the DPAC has been provided by national institutions, in particular the institutions participating in the {\it Gaia} Multilateral Agreement.
This study is based on work supported by NASA under Grant no. NNX15AB78G issued through the Astrophysical Data Analysis Program by Science Mission Directorate. M.E., W.D.C., and P.J.M. were supported by NASA K2 Guest Observer grants NNX15AV58G, NNX16AE70G and NNX16AE58G to the University of Texas at Austin. C.B. acknowledges support from the Alfred P. Sloan Foundation. 
Much thanks to Tim Morton for help with \emph{vespa}.
 
{\it Facilities:} \facility{\ktwo} \facility{Keck:II(NIRC2)}  \facility{KPNO:2.1m (Robo-AO)}.

\software{Batman 2.1.0 \url{http://astro.uchicago.edu/\~kreidberg/batman/}, pymodelfit 0.1.2 \url{https://pythonhosted.org/PyModelFit/core/pymodelfit.core.FunctionModel1D.html}, PyMC 2.3.4 \url{https://pymc-devs.github.io/pymc/}, Kea (Endl \& Cochran 2016), vespa 0.4.9 \url{https://github.com/timothydmorton/VESPA}, Uncertainties 2.4.6.1 \url{http://pythonhosted.org/uncertainties/)}}

\bibliography{k2_superpig_paperII}

\clearpage

\end{document}